\documentclass[conference]{IEEEtran}
\usepackage{amssymb}
\usepackage{amsmath}
\usepackage{cite}
\usepackage{verbatim}
\usepackage{textcomp}
\usepackage[pdftex]{graphicx}
\usepackage{grffile}
\usepackage{caption}
\usepackage{subcaption}
\usepackage{epsfig}
\usepackage{epstopdf}
\usepackage{caption}
\usepackage[ruled,vlined,linesnumbered, commentsnumbered]{algorithm2e}
\usepackage{balance}
\usepackage{listings}

\SetCommentSty{mycommfont}

\usepackage{tipa }	
\usepackage{pifont}	
\usepackage{wasysym}
\usepackage{color}

\usepackage{tikz}

\usepackage{setspace}

\usepackage{amsthm}

\DeclareMathOperator*{\argmin}{arg\,min}

\usepackage{color}

\begin{document}	

\title{Heterogeneity-aware P2P Wireless Energy Transfer for Balanced Energy Distribution}

\author{
\IEEEauthorblockN{Tamoghna Ojha, Theofanis P. Raptis, Marco Conti, and Andrea Passarella}
\IEEEauthorblockA{Institute for Informatics and Telematics, National Research Council, Italy\\
Email: \small \{tamoghna.ojha, theofanis.raptis, marco.conti, andrea.passarella\}@iit.cnr.it
}}


\maketitle

\begin{tikzpicture}[remember picture,overlay]
\node[anchor=south,yshift=10pt] at (current page.south) {\fbox{\parbox{\dimexpr\textwidth-\fboxsep-\fboxrule\relax}{
  \footnotesize{
    This work has been submitted to the IEEE for possible publication. Copyright may be transferred without notice, after which this version may no longer be accessible.
  }
}}};
\end{tikzpicture}

\begin{abstract}\label{Abst}
The recent advances in wireless energy transfer (WET) provide an alternate and reliable option for replenishing the battery of pervasive and portable devices, such as smartphones. The peer-to-peer (P2P) mode of WET brings improved flexibility to the charging process among the devices as they can maintain their mobility while replenishing their battery. Few existing works in P2P-WET unrealistically assume the nodes to be exchanging energy at every opportunity with any other node. Also, energy exchange between the nodes is not bounded by the energy transfer limit in that inter-node meeting duration. In this regard, the parametric heterogeneity (in terms of device's battery capacity and WET hardware) among the nodes also affects the energy transfer bound in each P2P interaction, and thus, may lead to unbalanced network energy distributions. This inherent heterogeneity aspect has not been adequately covered in the P2P-WET literature so far, especially from the point of view of maintaining a balanced energy distribution in the networked population. In this work, we present a Heterogeneity-aware Wireless Energy Transfer (HetWET) method. In contrast to the existing literature, we devise a fine-grained model of wireless energy transfer while considering the parametric heterogeneity of the participating devices. Thereafter, we enable the nodes to explore and dynamically decide the peers for energy exchange. The performance of HetWET is evaluated using extensive simulations with varying heterogeneity settings. The evaluation results demonstrate that HetWET can maintain lower energy losses and achieve more balanced energy variation distance compared to three different state-of-the-art methods.
\end{abstract}



\section{Introduction}\label{Intro}
Recent advancements in pervasive and portable device (such as smartphones) technologies have improved its available battery capacity significantly. However, with the increased user engagement, dependence and growing number of applications, the demand of device's energy resources has also increased significantly. Thus, the available energy remains limited for the device's continued operation, and the need for battery replenishment becomes obvious. In this regard, the recent advances in wireless energy transfer (WET) provides an alternate and reliable option for replenishing the battery of the smartphones \cite{Dhungana2020} \cite{Zhang2018}. WET has already matured through the development of numerous commercial products and standards, such as the ones of Wireless Power Consortium, Power Matters Alliance, Alliance for Wireless Power and Rezence \cite{Clerckx2022}. It enables the devices to use wireless charging from devices such as wireless charging pads and mobile charging vehicles \cite{Gao2020}, \cite{Li2019}, and even other peer devices \cite{PowerShare, huawei_wireless_charging}. This peer-to-peer (P2P) mode of WET brings improved flexibility in the charging process between smartphones as the devices are able to maintain their mobility while replenishing their battery. 

With the introduction of P2P-WET in latest smartphones, newer applications such as crowd charging \cite{Raptis2020} have emerged. Crowd charging facilitates distributed energy exchange throughout the network with simultaneous P2P interaction between multiple peers. In P2P-WET-based crowd charging applications, it is important to reach an `energy balanced' state, during which all the nodes reach an equal energy level. Thus, the energy balancing process also helps in extending the functional lifetime of the network as it replenishes the battery of nodes which have depleted their energy. However, the energy balance process is challenged by various factors -- the uncertainty of energy exchange due to opportunistic and varying inter-node meeting duration, and the energy loss associated with each P2P-WET interaction (due to wireless attenuation, energy loss in each WET interaction is inevitable). As a result, after a not-so-careful energy balancing strategy, the networked population may have an unbalanced energy distribution. 

In this regard, another important factor which challenges the energy balance process is the heterogeneity present in various parameters of the pervasive devices throughout the network \cite{Galinina2018}. Typically, the devices may have different values for battery capacity and WET hardware, which determines the Qi charging capacity (Qi the standard for WET developed by Wireless Power Consortium \cite{qi_wpc}). Due to such heterogeneity, the charging rates for P2P interaction between different combination of nodes will be different and result in varying amount of energy exchange between the devices. For example, the charging rate will be higher when the energy receiving device has lower battery capacity or the energy transmitting device has higher charging capacity. The opposite is also true vice versa.

The early works for energy balance in P2P-WET considers few unrealistic assumptions such as -- the nodes to be exchanging energy at each opportunity \cite{Nikoletseas2017} as well as with any other node \cite{Dhungana2019}. Also, in these works, the energy exchanged in any P2P interaction is not bounded by the possible energy that can be transferred in the inter-meeting duration of those nodes \cite{Dhungana2019, Dhungana2021}. Additionally, due to the heterogeneity present in the network, the varying amount of energy exchanged in different P2P interactions may lead to unbalanced energy distribution across the crowd as well as higher energy loss. The existing P2P-WET techniques did not consider the factor of heterogeneity present among the devices of the network. Therefore, the energy balancing process in those works is likely to be affected by the varying energy loss and unbalanced energy distribution.

\subsubsection{Novelty of the Current Work}
In this paper, we present a \underline{Het}erogeneity-aware \underline{W}ireless \underline{E}nergy \underline{T}ransfer  method, named \textit{HetWET}, for energy balancing of the network. Our objective in \textit{HetWET} is to achieve energy balancing of the network such that the overall energy loss is minimized and the total energy variation between the nodes also minimized. To achieve these objectives, \textit{HetWET} (a) first dynamically computes the charging rate for each P2P interaction, and (b) proactively estimates the energy loss and the variation in energy distribution for all possible peers. Then (c) the peers are chosen for P2P-WET such that the energy loss is minimized and distribution is balanced. In summary, our specific contributions in this work are the following:
\begin{itemize}
\item We devise a fine-grained model of network-wide P2P wireless energy transfer while considering the heterogeneity of the participating devices.
\item We design a method enables the nodes to explore and dynamically decide the peers for energy exchange.
\item We leverage the proactive estimation of the energy loss and energy variation for peer selection, and thereby minimizing both overall energy loss and variation.
\end{itemize}

The rest of the paper is organized as follows. In Section \ref{sec:RelWrk}, we discuss the related literature. The proposed network model and key concepts are presented in Section \ref{sec:SysModel}. The proposed \textit{HetWET} method is presented in Section \ref{sec:HetWET}, and in Section \ref{sec:Result}, we discuss the parameter evaluation settings, benchmarks and simulation results. Finally, Section \ref{sec:Conclu} concludes the paper citing directions for future research.

\section{Related Works}\label{sec:RelWrk}
In this section, we discuss the related works in two categories. The works in the first category focuses on energy balancing with low energy loss. Nikoletseas \textit{et al.} \cite{Nikoletseas2017} considered both loss-less and lossy WET, and then, estimate the upper bound of time required to energy balance the whole network for both the scenarios. The authors propose three methods for both the scenarios with objectives targeting minimization of energy loss and time to reach energy balance. In another work, Dhungana \textit{et al.} \cite{Dhungana2019} proposed three different energy exchange methods with the objective of minimizing the energy loss in the energy balance process. In contrast to the works of this category, in our proposed method, we consider a heterogeneous setting. We perform peer selection such that to minimize the energy loss as well as energy variation among the nodes while dynamically computing the energy to be exchanged between heterogeneous nodes.

The next category of works, focuses on different issues impacting the energy balance process. The impact of social networks on peer selection for P2P-WET was first introduced by Raptis \cite{Raptis2020}. The author devises two socially-motivated energy exchange strategies for the users explicitly focused on users' social relations. Ojha \textit{et al.} \cite{Ojha2021} highlighted the issues of user mobility affecting the P2P energy exchange. The authors leverage the mobility information for improving the peer selection process during P2P-WET. In another work \cite{Ojha2022a}, the authors incorporated the joint issue of user mobility as well as the influence of user's social impact on energy balancing process. The authors considered two dimensions of social information, social context and social relations, for predicting the P2P energy exchange opportunities. In \textit{HetWET}, we explicitly focus on the issue of device heterogeneity impacting the energy balance process and devise fine-grained model for heterogeneity-aware energy exchange for improved decision on peer selection.

\section{Network Model}\label{sec:SysModel}
We assume total $m$ users (also termed as `nodes' in this paper) each with a smartphone present in the network which is deployed over an area of interest. We denote the users as $\mathcal{U} = \{ u_1, u_2, \cdots, u_m \}$. We denote the residual energy levels of the users' devices at time $t$ as, $\mathcal{E}_t = \{ E_t(1), E_t(2), \cdots, E_t(m) \}$. The nodes are mobile and thus, their location changes over time. For example, at time $t$, the locations of the nodes are referred as $\mathtt{L}(t) = \{l^1_t, l^2_t, \cdots , l^n_t\}$. 

Unlike previous works, in this work, we incorporate the issue of heterogeneity among the users participating in P2P-WET. Subsequently, we assume that the users are equipped with different types of smartphones, different battery capacity and heterogeneous WET hardware. Now-a-days, the feature of P2P-WET is available in various smartphones, mostly high-end, by leading manufacturers such as Samsung and Huawei using the technology PowerShare \cite{PowerShare} and reverse wireless charging \cite{huawei_wireless_charging}. Using these technologies, one user can wirelessly charge in a P2P fashion another user's smartphone, assuming that both the devices are Qi certified \cite{qi_wpc} (compatible with Qi standard), and thus, compatible for P2P-WET. Let, $C_j$ and $V_j$ refer to the battery capacity ($mAh$) and voltage ($volts$) for user $u_j$. $QI_i$ denotes the charging capacity ($Wh$) of $u_i$'s smartphone when using PowerShare or reverse wireless charging technology. Next, we compute the charging rate ($\alpha_{ij}$) for any P2P-WET interaction ($u_i$ is the energy transmitter and $u_j$ is the receiver) where the devices have different battery capacity and WET hardware.

\begin{equation}
\alpha_{ij} = \frac{100}{60 \times \frac{C_j \times V_j}{1000 \times QI_i} }
\label{eq:alpha}
\end{equation} 
Here, the charging rate is computed in \% of charge/minute for any P2P-WET between $u_i$ and $u_j$. The values 100, 60, and 1000 are used to denote the full charge (100\%), hour to minute conversion, and $mWh$ to $Wh$ conversion of the denominator, respectively.

Now that we have computed the charging rate, we can calculate the amount of energy (or charge) transferred between $u_i$ and $u_j$ for a duration of $\tau_t^{ij}$. 
\begin{equation}
e_{ij}^t = \alpha_{ij} \times \tau_t^{ij}
\label{eq:e_ij}
\end{equation}

For a transfer of $e_{ij}^t$ energy from node $u_i$ to $u_j$ for a time duration ($\tau_t^{ij}$), $u_j$ receives only a fraction ($(1 - \beta) \times e_{ij}^t$) of the actual energy transferred by $u_i$ due to energy loss. Thus, the remaining energy of these nodes are computed as, $\big( E_t(i) - e_{ij}^t, E_t(j)+(1-\beta) e_{ij}^t \big)$, where $\beta \in [0,1)$ is the energy loss factor. Here, following the modeling approach of numerous works (such as \cite{Dhungana2019, Raptis2020, Bulut2020social}), we consider that $\beta$ is constant for all P2P-WET interactions between different heterogeneous devices. Also, the energy levels of any other node ($\forall u_k \in \mathcal{U}, u_k \neq u_i, u_j$) in the network remains unchanged for the energy transfer between nodes $u_i$ and $u_j$.

Next, we define a parameter named the energy variation distance which refers to the amount of energy variation among the deployed nodes. The computation of the variation distance is presented in \cite{ Nikoletseas2017, Dhungana2021} using probability theory and stochastic processes. For example, if $P$ and $Q$ are two probability distributions defined over the sample space of $\mathcal{U}$, then, the total variation distance, $\delta(P,Q)$, is computed as,
\begin{equation}
\delta(P,Q) = \sum_{x \in \mathcal{U}} |P(x) - Q(x)|
\label{eq:var_dist}
\end{equation}  

The energy distribution of nodes ($\mathcal{E}_t(u)$) at time $t$ is defined as,
\begin{equation}
\mathcal{E}_t(u) = \frac{E_t(u)}{E_t(\mathcal{U})} \qquad \mbox{ where, } E_t(\mathcal{U}) = \sum_{u \in \mathcal{U}} E_t(u)
\end{equation}
Subsequently, the average network energy is calculated as,
\begin{equation}
\overline{E}_t = \frac{E_t(\mathcal{U})}{m}.
\end{equation}

We assume that the nodes move from one location to another. The user movement in the considered area depends on their own interests only and not affected by the activity of other users. It is also assumed that the users tend to stay similar amount of time while revisiting any specific location. The users staying at same location for a certain amount of time can engage in P2P energy exchange. Specifically, the contact ($\nu_{ij}^t$) between any two nodes $u_i$ and $u_j$ for a duration of $\tau_t^{ij}$ is considered valid when the following conditions are satisfied $\forall t \in \tau_t^{ij}$,
\begin{equation} 
\nu_{ij}^t = 
\begin{cases}
1, \quad l^i_t = l^j_t \mbox{ and } \tau_t^{ij} \geq t_{min},\\
0, \quad \mbox{otherwise}
\end{cases}
\end{equation}
where $l^i_t$ and $l^j_t$ are the locations of $u_i$ and $u_j$. $t_{min}$ is the minimum required time for a successful P2P energy transfer.

Our objective in this paper is to minimize the overall energy loss (which maximizes the total network energy) and energy variation distance. Let, at $T$ time, the total energy loss is $\mathcal{L}_T$ and the energy variation distance is $\delta(\mathcal{E}_T, \mathcal{U}_T)$, where $\mathcal{U}_T$ is the target uniform energy distribution. 

The total energy of the network at time $T$, $E_T = \sum_{\forall u_i \in \mathcal{U}} E_T(i)$, is computed as,
\begin{equation}
\sum_{\forall u_i \in \mathcal{U}} E_T(i) = E_0(i) - \sum_{t \in T} \sum_{\forall u_j \neq u_i} e_{ij}^t + \sum_{t \in T} \sum_{\forall u_j \neq u_i} (1 - \beta)e_{ji}^t
\end{equation}
where $E_0(i)$ refers to the initial energy level of $u_i$. $\sum_{t \in T} \sum_{\forall u_j \neq u_i} e_{ij}^t$ and $\sum_{t \in T} \sum_{\forall u_j \neq u_i} (1 - \beta)e_{ji}^t$ compute the energy transmitted and received by $u_i$ from other nodes, respectively. 


\section{HetWET: Heterogeneity-aware Wireless Energy Transfer}\label{sec:HetWET}
In this section, we discuss the proposed heterogeneity-aware wireless energy balancing algorithm. We first define the parameters towards achieving our objective of minimizing energy loss and the variation distance. Thereafter, we present the algorithm showing the peer selection process for any user. 

First, we compute the energy loss during the P2P energy exchange between $u_i$ and $u_j$. It is computed as,
\begin{equation}
EL_{ij} = \beta \times e_{ij}^t
\label{eq:el_ij}
\end{equation}

Now, we compute the change in energy variation distance explicitly for the P2P-WET interaction between $u_i$ and $u_j$. The updated energy variation distance for $u_i$ and $u_j$ will be the difference between its energy level and average network energy. Therefore, the change in energy variation distance $EVD_{ij}$ is,
\begin{equation}
EVD_{ij} = \Big[ \Big( E_t(i) - e_{ij}^t \Big) - \overline{E}_t  \Big] + \Big[ \overline{E}_t - \Big( E_t(j)+(1-\beta) \times e_{ij}^t \Big) \Big]
\label{eq:evd_ij}
\end{equation}

It is important to note that both Equation \eqref{eq:el_ij} and \eqref{eq:evd_ij} are inclusive of the heterogeneity present between the nodes as we include the dynamic computation of the charging rate.

Next, we compute a factor named `selectivity factor' ($\Phi_{ij}$), which refers to the selection benefit for this pair of nodes ($u_i$ and $u_j$). $\Phi_{ij}$ considers both energy loss and energy variation distance for any P2P energy exchange between $u_i$ and $u_j$. As our objective is to minimize both energy loss and variation, we need to select the peers with lowest value of $\Phi_{ij}$. Here, we computed $\Phi_{ij}$ as,
\begin{equation}
\Phi_{ij} = w_{EL} \times \frac{EL_{ij}}{\sum_{u_k \in N_{u_i}} EL_{ik}} + w_{EVD} \times \frac{EVD_{ij}}{\sum_{u_k \in N_{u_i}} EVD_{ik}}
\label{eq:sel_fac}
\end{equation}
where, $w_{EL}$ and $w_{EVD}$ denote the weightage for the energy loss and energy variation distance components. $N_{u_i}$ denotes the set of possible peers of $u_i$. $N_{u_i}$ is computed as the set of users satisfying a valid contact, and their energy levels in the opposite side of energy balance level compared to $u_i$ . This design choice of choosing nodes with energy in opposite sides of energy balance level ensures higher possibility for both nodes reaching energy balance level. Interested readers may refer to \cite{Dhungana2019, Ojha2021}. Thus, the computation of $N_{u_i}$ is as follows,
\begin{multline}
N_{u_i} = \bigcup_{\forall u_k, l^{u_k}_t = l^{u_i}_t} u_k		\quad \mbox{where, } E_t(u_k) < \overline{E}_t \mbox{ if } E_t(u_i) > \overline{E}_t \\
\mbox{\textit{Or, }} E_t(u_k) > \overline{E}_t \mbox{ if } E_t(u_i) < \overline{E}_t
\end{multline}

Now, our objective for the P2P energy exchange over the whole network is to minimize the energy loss as well as the energy variation distance. Therefore, we just need to choose the user with minimum value of selectivity factor for energy exchange with $u_i$ such that both energy loss and energy variation distance remain minimum.
\begin{equation}
u_j = \argmin_{u_j \in \mathcal{U}} \Phi_{ij}		\qquad u_j \ne u_i
\end{equation}

In Algorithm \ref{algo:hetWET}, we list the steps followed for finding the peer for any node while considering heterogeneous configurations of the devices. We first (lines 5-6) select the node (say, $u_i$) with energy level closest to the target energy balance level ($\overline{E}^*$), as $u_i$ is the closest node to reach energy balance level. $\overline{E}^*$ is computed as shown by \cite{Dhungana2019},
\begin{equation}
\overline{E}^* = \frac{-(1-\beta) + \sqrt[2]{(1-\beta)}}{\beta} \qquad \forall \beta \in [0, 1], m \longrightarrow \infty
\end{equation}
Next, we find the potential peers ($N_{u_i}$) for $u_i$ (lines 7-9) and then, for interaction with each such peer, we estimate the energy loss and variation distance, and the selectivity factor (lines 10-17). We, then, find the peer (say, $u_j$) with minimum value of the selectivity factor as the selected peer for $u_i$ (line 18). Then, the P2P energy exchange is performed between each such pair of peers, and if the nodes' updated energy equals $\overline{E}^*$, we consider it as being reached energy balance level.

\begin{algorithm}[t!]
\caption{Heterogeneity-aware Peer Selection \label{algo:hetWET}}
\textbf{Inputs:} $\mathcal{E}_t$, $|C_i|_{\forall u_i \in \mathcal{U}}$, $|QI_i|_{\forall u_i \in \mathcal{U}}$.\\
\textbf{Output:} Peer selection.\\
\SetKwFunction{MyFunc}{ComputeForPotentialPeers}
Initialize $State[\cdot] \longleftarrow Incomplete$\;
\While{$t \leq T$}{
\For{$u_i \in \mathcal{U}$ and $State[i] = Incomplete$}{
	Find the node with energy closer to $\overline{E}^*$, $u_i \longleftarrow \argmin_{u_i \in \mathcal{U}} |\overline{E}^* - E_t(i)|$\;
}
\For{$u_j \in \mathcal{U}$ and $u_j \neq u_i$}{
	\If{$\nu_{ij}^t = 1$ and $State[j] = Incomplete$}{ 
		$N_{u_i} \longleftarrow N_{u_i} \cup u_j$\;	
	}
}
\eIf{$E_t(i) > \overline{E}^*$} { 
	\For{$u_j \in N_{u_i}$ and $E_t(j) < \overline{E}^*$}{
		ComputeForPotentialPeers($u_i$, $u_j$)\;
	}
	Compute $u_j = \argmin_{u_j \in \mathcal{U}} \Phi_{ij}$\;
}{ 
	\For{$u_j \in N_{u_i}$ and $E_t(j) > \overline{E}^*$}{
		ComputeForPotentialPeers($u_j$, $u_i$)\;
	}
	Compute $u_j = \argmin_{u_j \in \mathcal{U}} \Phi_{ij}$\;
}
Return $u_j$ as peer of $u_i$\;
}
\SetKwProg{Fn}{Function}{:}{}
\Fn{\MyFunc{$u_x$, $u_y$}}{
	Compute $\alpha_{xy}$ and $e_{xy}$ using Equations \eqref{eq:alpha} and \eqref{eq:e_ij}\;
	Compute $EL_{xy}$ using Equation \eqref{eq:el_ij}\;
	Compute $EVD_{xy}$ using Equation \eqref{eq:evd_ij}\;
	Compute $\Phi_{xy}$ using Equation \eqref{eq:sel_fac}\;
}
\end{algorithm}

\section{Performance Evaluation}\label{sec:Result}

\subsection{Simulation Settings}\label{sec:sim_settings}
In our experiments, we consider 100 smartphone carrying users distributed uniformly at random over 5 different locations. These smartphones have different battery capacity ($C_i$) and charging capacity ($QI_i$) distributed uniformly at random  allocated from three different set of values: $(C_i, QI_i) \in \{(5000, 5), (4500, 4), (4000, 3)\}$. In another set of experiments, we also vary the heterogeneity by allocating different number of nodes (rather than being uniform distribution). For example, in three different experiments, we allocate (50\%, 25\%, 25\%), (25\%, 50\%, 25\%) and (25\%, 25\%, 50\%) nodes for the three set of battery and charging capacity values, respectively. In another set of experiments, we vary heterogeneity by expanding the range of values for $C_i$ and $QI_i$. For example, we use the following formula for generating heterogeneous set of battery and charging capacity values: $(C_i, QI_i) \in \{(4500 + \theta \times \delta_{C}, 4 + \theta \times \delta_{Q}), (4500, 4), (4500 - \theta \times \delta_{C}, 4 - \theta \times \delta_{Q})\}$. Where, we consider $\delta_{C}$ and $\delta_{Q}$ as 500 and 1 (to generate close to realistic values of ($C_i, QI_i$), respectively, and vary $\theta$ with values 0.5, 1.5 and 2.0 in three different experimental settings, which generates uniform distribution of nodes with different ($C_i, QI_i$) values.

We perform our experiments over multiple `iterations' or `runs'. In each iteration of the experiment, the nodes randomly select a location and stay there for a randomly selected duration chosen over 10-30 $minutes$, and then move to another randomly selected location in the next iteration. Therefore, the concept of iteration actually works as the specific decision boundary for the P2P interactions. As energy exchange between any two peers is bounded by their inter-meeting duration, the concept of time is also enforced in the experiments. The initial energy distribution of the nodes are uniformly randomly distributed over $[0, 100]$ units. The charging rate $\alpha_{ij}$ is dynamically computed in each P2P interaction as shown in Equation \eqref{eq:alpha}, and the energy loss rate is considered as $\beta = 0.2$. We repeat each simulation experiment for 50 times for statistical smoothness.

\subsection{Benchmarks}
To compare the performance of the proposed method, \textit{HetWET}, we consider three state-of-the-art methods, namely \textit{MobiWEB} \cite{Ojha2021}, $P_{GO}$ \cite{Dhungana2019} and $P_{OA}$ \cite{Nikoletseas2017}, as the benchmarks. Both \textit{MobiWEB} and $P_{GO}$, select peers with energy in the opposite side as well as closest to the energy balance level. Next, the selected peers exchange energy among themselves such that the nodes closest to the energy balance level reaches it. However, \textit{MobiWEB} leverages the user mobility information for intelligent peer selection. Whereas, in $P_{OA}$, nodes with energy levels in the opposite sides of average energy, participate in energy exchange such that their energy levels become equal. To have a fair comparison with \textit{HetWET}, we made few adjustments in the benchmarks: the P2P energy exchange remains bounded by the corresponding charging rate ($\alpha_{ij}$) and inter-meeting time ($\tau_t^{ij}$), the nodes in $P_{OA}$ have knowledge of average network energy. 

We discuss the results of the proposed and benchmark methods using the following metrics: \textit{total network energy}, \textit{total energy variation distance}, \textit{number of P2P meetings}, and \textit{number of nodes that reached energy balance}. We also perform a sensitivity analysis and discuss the results of \textit{total energy variation distance} varying the \textit{heterogeneity} present in the network. We choose \textit{total energy variation distance} as this parameter shows the quality of energy balance in the resulting network.

\begin{figure*}[ht]
        \centering
        \begin{subfigure}[b]{0.25\textwidth}
                \includegraphics[width=\textwidth]{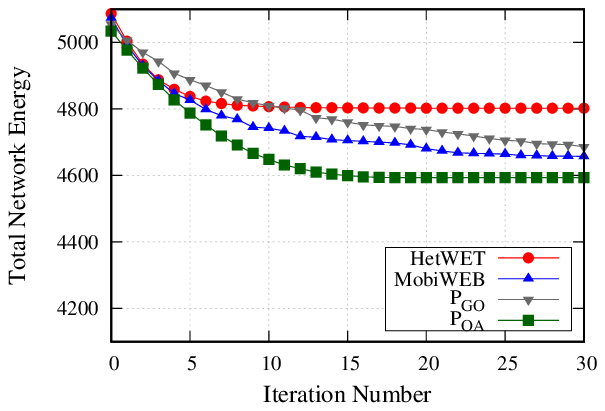}
                \caption{Total network energy.}
                \label{fig:tot_energy}
        \end{subfigure}%
        \begin{subfigure}[b]{0.25\textwidth}
                \includegraphics[width=\textwidth]{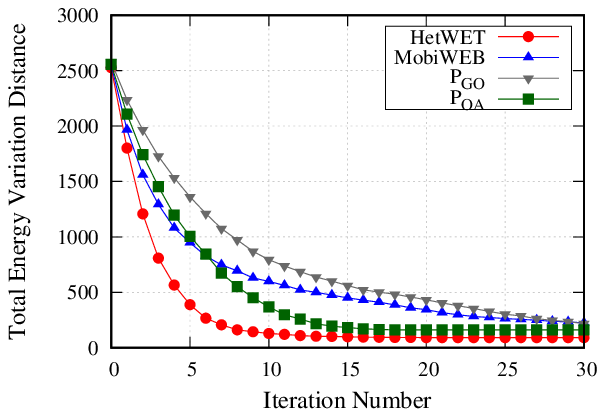}
                \caption{Total energy variation distance.}
                \label{fig:tot_var_dist}
        \end{subfigure}%
        \begin{subfigure}[b]{0.25\textwidth}
                \includegraphics[width=\textwidth]{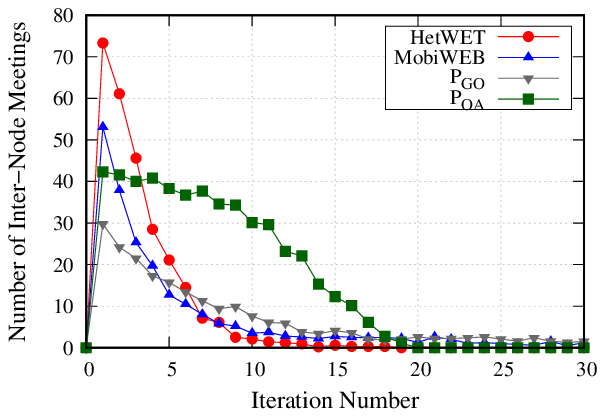}
                \caption{Number of P2P meetings.}
                \label{fig:num_meet}
        \end{subfigure}%
        \begin{subfigure}[b]{0.25\textwidth}
                \includegraphics[width=\textwidth]{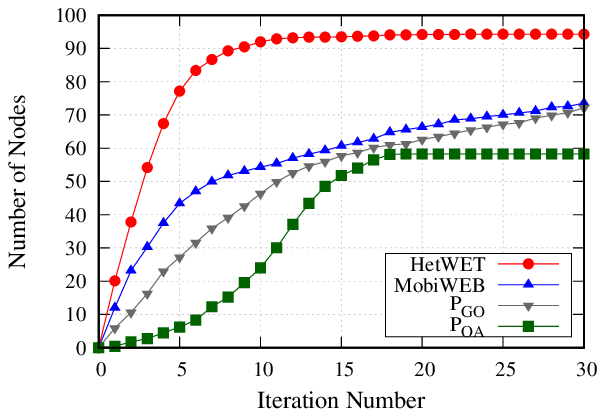}
                \caption{Nodes reached energy balance.}
                \label{fig:num_complete}
        \end{subfigure}%
        \caption{Performance comparison of \textit{HetWET} with other state-of-the-art methods.}
\end{figure*}

\subsection{Results}

\subsubsection{Total network energy}
In Figure \ref{fig:tot_energy}, we present the results for the total network energy for the proposed and benchmark methods. The results show that \textit{HetWET} is able to maintain an overall higher network energy over time. It is evident from the results that the proactive consideration of energy loss while selecting the peers from available heterogeneous set of potential peers in the proposed method has helped in maintaining lower energy loss in the whole network, and thus, the total network energy remains higher in \textit{HetWET} compared to the benchmarks. In \textit{MobiWEB} and $P_{GO}$, the peer selection is performed only based on their energy levels, and dynamic computation (required due to network's inherent heterogeneity) of possible energy that can be exchanged is not considered. On the other hand, in the $P_{OA}$ method, no specific criteria is allocated for peer selection. In the initial iterations (1-10), \textit{HetWET} shows higher energy loss due to higher number of P2P interactions performed in that duration, as shown in Figure \ref{fig:num_meet}.

\subsubsection{Total energy variation distance} 
Figure \ref{fig:tot_var_dist} shows the total energy variation distance in the resulting network for all the methods. The \textit{HetWET method} is able to outperform the benchmarks. Compared to the benchmarks, the proposed method leverages the estimated total variation distance information and applies it to select peers. Also, for any P2P interaction, the computation of energy loss as well as resulting variation distance is based on their corresponding battery capacity and charging capacity. Due to these two reasons \textit{HetWET} is able to maintain a lower energy variation distance compared to the benchmarks.

\subsubsection{Number of P2P meetings} 
The results for number of P2P interactions in each iteration is shown in Figure \ref{fig:num_meet}. In the initial iterations (0-5), all the methods show higher number of P2P interactions compared to the rest of the iterations. However, in \textit{HetWET}, the number of such interactions are higher compared to the benchmarks. Thus, it is evident that the proposed method is able to promote higher number of P2P meeting opportunities. The significant decrease in energy variation distance and energy loss is also due to this higher number of interactions. As shown in Figure \ref{fig:num_complete}, most number of nodes also reach energy balance in the early iterations (0-5).

\subsubsection{Number of nodes that reach energy balance} 
In Figure \ref{fig:num_complete}, we depict the number of nodes that reach energy balance for all the methods. The results show that in \textit{HetWET} more number of nodes reach energy balance compared to benchmarks. Such performance is attributed to the dynamic computation of possible energy exchange, and the proactive estimation of energy loss and variation distance. As we find that a higher number of the nodes reach energy balance while maintaining lower energy variation distance, we can infer that \textit{HetWET} provides improved energy balance quality compared to the benchmarks.

\begin{figure*}[ht]
        \centering
        \begin{subfigure}[b]{0.25\textwidth}
                \includegraphics[width=\textwidth]{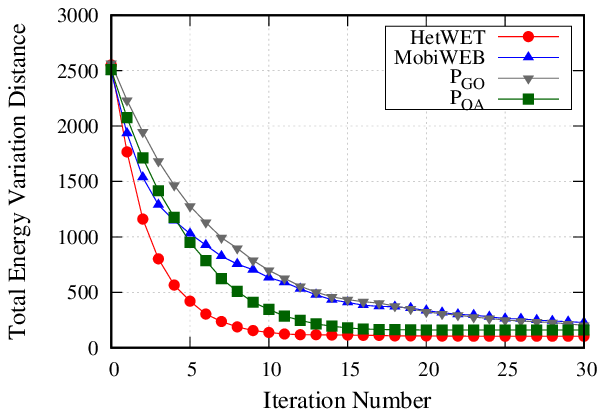}
                \caption{(50, 25, 25).}
                \label{fig:var_dist_50_25_25}
        \end{subfigure}%
        \begin{subfigure}[b]{0.25\textwidth}
                \includegraphics[width=\textwidth]{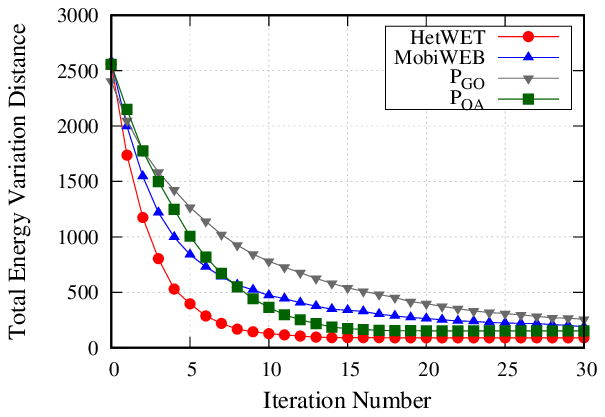}
                \caption{(25, 50, 25).}
                \label{fig:var_dist_25_50_25}
        \end{subfigure}%
        \begin{subfigure}[b]{0.25\textwidth}
                \includegraphics[width=\textwidth]{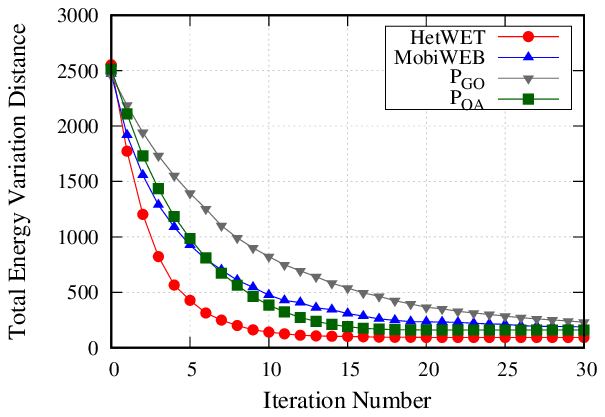}
                \caption{(25, 25, 50).}
                \label{fig:var_dist_25_25_50}
        \end{subfigure}%
        \caption{Comparison of total energy variation distance with different heterogeneous setting.}
\end{figure*}

\subsubsection{Effect of heterogeneity in the network}
\begin{figure*}[ht]
        \centering
        \begin{subfigure}[b]{0.25\textwidth}
                \includegraphics[width=\textwidth]{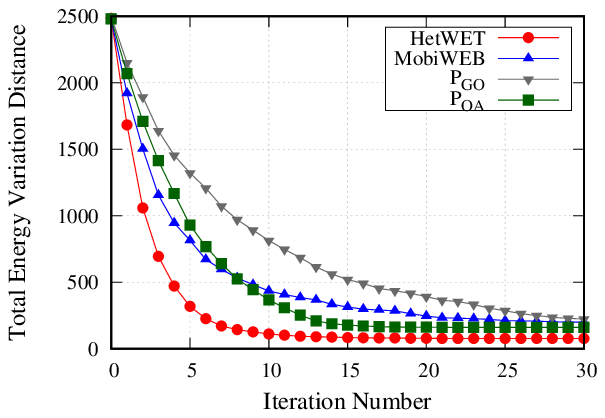}
                \caption{$\theta = 0.5$.}
                \label{fig:var_dist_0.5}
        \end{subfigure}%
        \begin{subfigure}[b]{0.25\textwidth}
                \includegraphics[width=\textwidth]{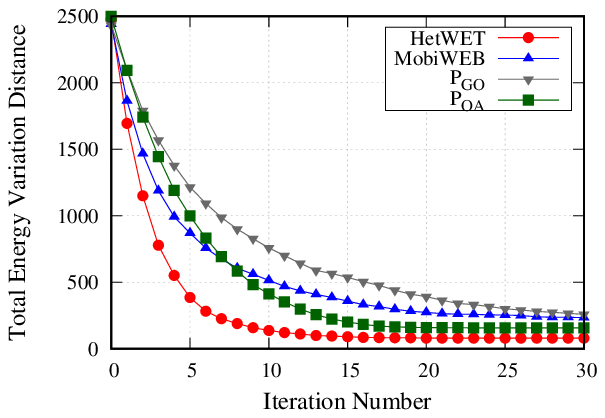}
                \caption{$\theta = 1.5$.}
                \label{fig:var_dist_1.5}
        \end{subfigure}%
        \begin{subfigure}[b]{0.25\textwidth}
                \includegraphics[width=\textwidth]{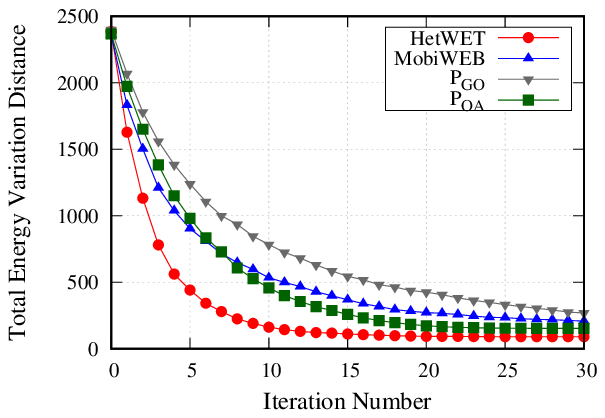}
                \caption{$\theta = 2.0$.}
                \label{fig:var_dist_2.0}
        \end{subfigure}%
        \caption{Comparison of total energy variation distance with different heterogeneous setting.}
\end{figure*}

until now, we discussed the performance evaluation for uniform distribution of battery and charging capacity across the nodes. In the following set of experiments, to vary heterogeneity among the network, we allocate (50\%, 25\%, 25\%), (25\%, 50\%, 25\%) and (25\%, 25\%, 50\%) nodes for the three set of battery and charging capacity values, respectively. We show the results for these set of experiments in Figures \ref{fig:var_dist_50_25_25} --\ref{fig:var_dist_25_25_50}. In another set of experiments, we vary the values of the battery and charging capacity while uniformly allocating the values to the nodes. In Section \ref{sec:sim_settings}, we discuss how these values are varied w.r.t. a parameter named $\theta$. In Figures \ref{fig:var_dist_0.5} -- \ref{fig:var_dist_2.0}, we show the results for these experiments for $\theta = 0.5$ to $\theta = 2.0$.

Both these sets of experiments show that the proposed method is able to maintain a lower total energy variation distance compared to the benchmarks. In both set of experiments, the benchmarks also perform nearly similarly with minor variation. Such behavior is attributed to the proactive estimation of energy variation distance for the P2P interactions and the dynamic selection leveraging this information. We see that $P_{OA}$ performs comparatively better than other two benchmarks w.r.t. energy variation distance. However, $P_{OA}$ also results in higher energy loss and lower number of nodes reaching energy balance level.

Therefore, overall with different heterogeneous settings, we can infer that the proposed method is able to maintain a better quality of energy balance in terms of lower energy loss and variation distance, and higher number of nodes reaching the energy balance level.

\section{Conclusion}\label{sec:Conclu}
In this paper, we consider the issue of heterogeneity present among the devices participating in P2P-WET affecting the energy balance process. Specifically, we consider the heterogeneity in terms of the participating device's battery capacity and WET hardware, which determines the Qi charging capacity. We propose a method, Heterogeneity-aware Wireless Energy Transfer (HetWET), which includes a dynamic and fine-grained model of energy exchange. Our proposed method enables the nodes to explore the potential peers, and subsequently, select a peer dynamically for P2P energy exchange. Compared to the existing works, HetWET achieves improved energy balance quality w.r.t different performance metrics. Also, with varying heterogeneous settings, the proposed method is able to maintain low energy loss and energy variation distance compared to the state-of-the-art. In future, we plan to extend our work considering additional energy losses (due to different activities e.g. mobility, communication) and implement in a real-world application scenario.

\section*{Acknowledgment}
This work was carried out during the tenure of an ERCIM ‘Alain Bensoussan' Fellowship Programme of the first author.

\bibliographystyle{IEEEtran}
\bibliography{main}

\balance

\end{document}